# APPLICATION OF MACHINE LEARNING FOR INFRASTRUCTURE RECONSTRUCTION PROGRAMS MANAGEMENT

*Illia Khudiakov, Vladyslav Pliuhin, Sergiy Plankovskyy, Yevgen Tsegelnyk*
*O.M. Beketov National University of Urban Economy in Kharkiv, Kharkiv, Ukraine*
*Illya.Hudyakov@kname.edu.ua*

**Abstract** – The purpose of this article is to describe an adaptive decision-making support model aimed at improving the efficiency of engineering infrastructure reconstruction program management in the context of developing the architecture and work breakdown structure of programs. As part of the study, the existing adaptive program management tools are analyzed, the use of infrastructure systems modelling tools is justified for program architecture and WBS creation. Existing models and modelling methods are viewed, and machine learning and artificial neural networks are selected for the model. The main components of the model are defined, which include a set of decision-maker preferences, decision-making tasks, sets of input data, and applied software components of the model. To support decision-making, the adaptive model applies the method of system modeling and predicting the value of the objective function at a given system configuration. Prediction is done using machine learning methods based on a dataset consisting of historical data related to existing engineering systems. The work describes the components of the redistribution of varied model parameters, which modify the model dataset based on the selected object type, which allows adapting the decision-making process to the existing program implementation goals. The functional composition done in Microsoft Azure Machine Learning Studio is described. The neural network parameters and evaluation results are given. The application of the developed adaptive model is possible in the management of programs for the reconstruction of such engineering systems as systems of heat, gas, electricity supply, water supply, and drainage, etc.

**Keywords**: neural network, machine learning, program management, decision-making support, infrastructure reconstruction, Microsoft Azure Machine Learning Studio

## 1. Introduction

The implementation of Engineering Infrastructure Reconstruction Programs (EIRP) in urban areas is vital for both the efficient operation of communal systems and the broader reconstruction of Ukraine following the full-scale invasion by the Russian Federation.

Within EIRP management, particular emphasis is placed on designing the program's architecture and structuring project hierarchies during the planning phase, as well as managing the architecture throughout implementation. A critical aspect of this process involves making informed management decisions about selecting equipment for installation at reconstructed sites. These decisions must consider the current energy demands of consumers, compliance with state regulations, and the specifications of existing equipment at system facilities that are not slated for replacement under the program.

The constraints of infrastructure programs and the impact of a volatile external environment highlight the necessity of employing adaptive management methods in overseeing private investment projects. This approach is particularly relevant when making management decisions related to program architecture, as well as in developing and managing the hierarchical structure of project tasks.

The topic of adaptive management, both in general and in the specific context of programs and projects, has been extensively explored by numerous researchers [1-5]. Scholars have proposed adaptive management approaches for programs [6], applicable across various implementation phases.

Adaptive decision support systems have been examined in studies [7-9]. However, the use of such tools in project and program management processes remains underexplored. Notable works include [10], which proposed an adaptive project management model for developing a professional doctorate in business management. Despite this, there are no current studies addressing the application of adaptive decision support tools in engineering infrastructure reconstruction programs.

Work [11] investigates internal and external variables influencing large program management, proposing a system dynamics model that incorporates these factors to enhance program risk management. For engineering infrastructure reconstruction programs, a methodology has been developed [12] for probabilistic quantitative modeling of individual restoration steps, considering environmental and human factors.

In managing engineering infrastructure reconstruction programs, engineering systems modeling and optimization tools can support decision-making. However, no existing tool effectively address decision-making needs for program architecture development, hierarchical task structuring, or program management processes. While systems simulation tools [13-17] offer limited applicability, they exhibit several significant limitations:

- Insufficient representation of system complexity, including parameters and interrelations between elements.
- Lack of actionable decision-making guidance for program managers.
- Inability to scale for application across different system levels.
- Usability challenges for decision-makers.

Other modeling methods should be considered. A comparative analysis is presented in Table 1.

*Table 1. Comparative analysis of modeling methods*

| Modelling method | Advantages | Disadvantages |
|---|---|---|
| Models based on genetic algorithms | 1. Taking into account non-linear relationships between the target indicator and input data; 2. The ability to determine optimal combinations of characteristics of the modeled object. | 1. The need to use only quantitative data; 2. Wide spread of calculation results relative to the absolute extreme. |
| Cartesian product method | 1. High accuracy of calculations; 2. The ability to take into account dependencies between the characteristics of the modeled object. | 1. The need to use only quantitative data; 2. The need to use large computing power to calculate the target indicator. |
| Regression models | 1. High accuracy of calculations; 2. The ability to determine the degree of importance of | 1. The need to use only quantitative data. |
| | variables thanks to coefficients. | |

Thus, taking into account the shortcomings of the above models, there is a need to create a tool that will be based on the following principles:

- Complexity – elements of infrastructure systems have a wide range of parameters that are not limited to their technical characteristics; accordingly, the tool should apply a comprehensive approach to modeling and also take into account environmental, economic parameters of system elements, etc.;
- Simplicity and convenience – the tool should be accessible to the person making management decisions when implementing programs and projects for the reconstruction of engineering infrastructure, without the need to use software products that require the user to have special competencies and knowledge;
- Applicability in management processes – the modeling results should carry elements of hints for the decision-maker that can be used when making management decisions;
- Scalability – the tool should have the potential to scale the modeled systems if necessary;
- Possibilities of applying data categories – the tool should provide for the use of both quantitative and qualitative data.

It is necessary to ensure the ability to adapt program management processes to internal and external influences, therefore adaptability is highlighted as an additional principle of developing a decision support tool. Decision support should be implemented both when developing program architecture by selecting projects that will be part of them, and when managing the architecture during the program implementation process.

Using machine learning technologies should be used to develop the tool [18-22], since their application allows to ensure its compliance with the above criteria. Among machine learning methods, artificial neural networks are more effective for the goals of the research [23-26] since models based on them most perfectly implement the process of processing data categories, which allows to ensure the processing of both quantitative and qualitative features of system objects by the model.

This paper proposes a decision-support model based on machine learning methods, namely neural network regression, allowing for the appropriate selection of the scope of projects within EIRP architecture. This is achieved by selecting equipment composition for reconstruction projects by predicting carbon dioxide emissions value as a target function for different equipment sets. Such a model allows for taking into consideration the internal and external influences on the program and adapting the program architecture correspondingly through the selection of the characteristics of the reconstruction object, while considering the environmental impact of the resulting system.

## 2. Research Methodology

Decision support using the proposed model relies on simulating the system-object undergoing reconstruction. Engineering infrastructure systems are inherently complex organizational and technical structures, making their formalization through conventional methods, such as mathematical modeling, challenging to implement [27, 28]. Therefore, it is more practical to represent them as an array of elements with predefined parameters.

To develop such a model, an artificial neural network is employed. This approach eliminates the need for the decision-maker to manually calculate the objective function for given input values, as predictions are generated autonomously based on historical data, without requiring user intervention.

Due to the presence of a potential impact on the systems-objects of modeling of the turbulent external environment, the impact of the internal environment, the discount forecasting method was used to develop the model. Regressive [29] and autoregressive models [30] well reflect stable trends of the predicted indicators. In order for the models to reflect trends well, it is necessary to use a selective statistical set of parameters of the model parameter matrix X and the vector of predicted values of the target indicator in a sufficiently large sample dimension $N$ so that $N >> n$, where $n$ is the number of independent parameters.

However, in this case, it is possible to transfer to the forecasting stage some trends of changes in the predicted indicators, characteristic of distant time intervals of the past and not characteristic of the nearest and predicted time intervals. In order to reduce the specified disadvantage of trend models without reducing the sample dimension, a discount forecasting model is used.

Using artificial neural networks for predicting the target value was chosen due to their flexibility, adaptability, and forecasting accuracy. It is also possible to use them for small data samples, which is advisable when using them to train data models related to engineering infrastructure objects due to restrictions on access to such data. In addition, artificial neural networks allow the use of both numerical and qualitative data, which allows to increase the forecasting accuracy due to the use of a wider range of input parameters.

The mathematical model of a neural network is described by the expression:

$$Y = m_{CO} = \sum_{i=1}^{m} \omega_{jk} y_i \qquad (1)$$

Here, $y_i$ represents layer outputs, $\omega_{jk}$ represents weight coefficients of the j-th neuron of the k-th layer. Objective function $Y$ equals specific $CO_2$ emissions for the system $m_{CO}$. Using $m_{CO}$ as an objective function allows to include the environmental component in the decision-making process focusing also on greenhouse gasses emission decrease as a result of infrastructure system reconstruction. In the described model, mCO being the target function, is the output value of the neural network representing the parameter on the basis of which the decision is made.

$$y_i = \phi(net_i) = \phi\left(\sum_{i=1}^{m} \omega_{ij} X_i\right) \qquad (2)$$

$\phi(net_i)$ represents the hidden layer activation function, $X_i$ represents input data; $X_i$ represents the set of input data.

According to the least squares method, the network operation error is determined by expression (3), where $D_i$ represents actual values of the output function.

$$\sum E_p = \frac{1}{2} \sum_{i=1}^{m} (Y_i - D_i)^2 \qquad (3)$$

In the case considered in this paper, decision-making consists in choosing such a composition and configuration of the system from among the available alternatives, at which the minimum value of the objective function $m_{CO}$ will be achieved, by predicting its values at different values of the parameters of the system elements to comply with the program constraints. The guiding influence is expressed in the selection of projects for the program architecture or the development of the content and hierarchical decomposition of the work of projects related to the replacement or installation of equipment at the facilities of the infrastructure system. Thus, $m_{CO}$ is a minimization function for the input parameters $x_1, x_2, ..., x_n$:

$$m_{CO} = \min (x_1, x_2, ..., x_n) \qquad (4)$$

When there is a need to exercise managerial influence, the decision-maker determines the system parameters and existing restrictions, sends the corresponding request to the model. The request data is processed by a neural network, the value of the objective function is predicted, after which the result is transmitted to the user interface.

It is worth noting that making changes to the architecture of the engineering infrastructure reconstruction program is advisable only in the presence of external or internal influences that cause changes in the values of the parameters of the system elements. In this case, the model can also be used as a tool for predicting the results of possible managerial influences and environmental influences as one of the elements of an adaptive approach to program management.

The structure of the decision support model in general includes the set of decision maker's advantages $B$, the set of input data $I$, the task of predicting the objective function value $T_F$, the set of application software components $P_A$:

$$DSM = < B, I, T_F, P_A >  \qquad (5)$$

The set of decision maker's advantages includes program or project constraints, personal experience, and expertise.

The set of input data includes the object type $I_{obj}$ and the set of model limitations $I_{lim}$, $I_{lim} = \vec{X}$ where $\vec{X}$ is the input data vector, consisting of the set of input data $X_i$.

$$I = \{I_{obj}, I_{lim}\} \qquad (6)$$

$I_{obj}$ is a qualitative variable, the value of which determines the composition of model constraints used to predict the value of the objective function. Infrastructure objects of different types are characterized by differences in purpose and structural composition, which implies the presence of unique parameters for different types of objects. Thus, $I_{obj}$ acts as a parameter that determines the composition of parameters $\vec{X}$ for a specific system. $I_{obj}$ is a manually designed parameter, defined based on user input when selecting the type of object for prediction.

The model constraints are the values of the parameters of the elements of the object to be reconstructed, on the basis of which the value of the objective function is predicted. The type of object as an element of input data allows for the redistribution of parameters for forecasting, depending on the purpose of exercising control influence. Thus, when managing a program for the reconstruction of engineering systems, it is possible to make decisions within the framework of the development and management of architecture by selecting projects for the reconstruction of subsystems or individual objects of subsystems with such characteristics that would satisfy the general program constraints.

The model for predicting the values of system parameters includes the artificial neural network *ANN*, the dataset *DS*, the variable parameters redistribution components *VPRC* and the post-processing mechanism *PPM*:

$$F = \{ANN, DS, VPRC, PPM\} \qquad (6)$$

The varied parameters redistribution components are elements of the model that adapt it to the goals of modeling.

The purpose of the program for the reconstruction of engineering infrastructure is the reconstruction of the system as a whole, one or more subsystems. Within the framework of the management of such programs, there is a need to make decisions regarding the selection of projects for the reconstruction of objects that are part of the subsystems. Such objects, being part of one subsystem, may have technical and technological differences, which require an individual approach to supporting decision-making on their reconstruction. Examples of such objects are boiler houses and CHPs in heat supply systems, condensing, nuclear, and hydroelectric power plants in power supply systems, etc. Therefore, it is necessary to ensure the possibility of adapting the decision-making process to the presence of such objects in the system to be reconstructed within the framework of the program. When implementing projects for the reconstruction of individual objects, outside the program, there is also a need to form such a set of model constraints that will correspond to the constraints of the corresponding type of object.

The aforementioned components adjust and redistribute parameters for various types of subsystem objects.

The redistribution process is guided by the value of the linguistic variable $I_{obj}$. This involves filtering the dataset to exclude parameters that are unique to object types other than the selected one and are not shared across all object types. The result is a refined dataset that is specific to the chosen object type.

The data post-processing mechanism enhances decision-making by enabling consideration of the interrelationships among system elements during parameter optimization. It provides visual insights into the influence of key parameters on the target indicator, presenting this information as graphs that illustrate indicator dependencies.

The scheme of data post-processing is shown in Figure 1.

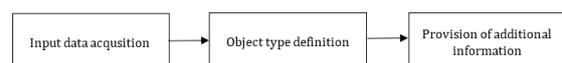

*Figure 1: Data post-processing scheme*

The scheme of the decision-making process in the management of the EIRP in the Business Process Model and Notation (BPMN) notation is shown in figure 2.

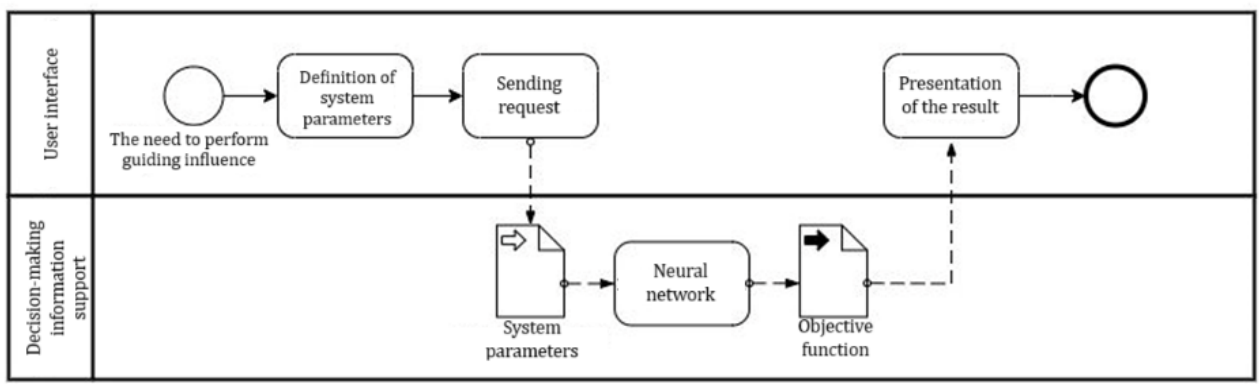

*Figure 2: Diagram of the decision-making process when managing engineering infrastructure reconstruction programs in BPMN notation*

The model is both a means of modeling the management object and a means of predicting the results of implementing management influences. Given this, the result of using the model in the program management processes will be a model of the system to be reconstructed within the framework of the implementation of the EIRP, and the forecast values of the objective function used for decision-making in the management of the EIRP.

## 3. Functional composition of the model

The decision support model was developed using the Microsoft Azure Machine Learning Studio. The use of this service is due to the following factors:

- Ease of use – the components of the machine learning system are represented by blocks that have inputs and outputs, respectively, the output data of one block is input for another; thus, there is no need to write code to create the model;
- Wide functionality – The Studio has a large number of forecasting and data classification methods for training the model;
- The ability to add user scripts to the model – additional data processing and analysis tools are implemented using scripts in the Python and R programming languages;
- Convenience of analyzing the quality of model training – The Studio provides the ability to analyze the quality of the model using a number of metrics: root-mean-square error, relative absolute error, coefficient of determination, etc.

Defining model limitations is a user function, and it is possible to define constraints for both quantitative and qualitative indicators. Thus, for an existing system that is subject to reconstruction within the framework of the implemented program, it is possible to adapt individual model parameters to the existing environmental disturbances.

The model operation involves specifying all parameter values that act as constraints. This is due to the limitations of the machine learning technology used and the Microsoft Azure Machine Learning Studio.

Two separate models were created for each type of object. These models are identical in architecture and differ in the content of two of the available blocks.

The input data for training the model is provided in the form of a dataset created in CSV format.

The dataset used for training of the model includes 56 columns and 100 rows of data for the cogeneration plants and boiler houses in Ukraine gathered from the open sources for 2018-2022. The rows including the data on objects of different types in the initial dataset were given together being later separated by custom Python script described below.

The data is separated into 6 logical blocks, including the following:

- technical and technological parameters: main fuel type, total heat power, number of pumps, annual specific fuel consumption for heat energy generation etc.;
- environmental parameters: annual $CO$ emissions, annual $SO$ emissions, type of fuel etc.;
- energy efficiency parameters: pump energy efficiency class, specific annual fuel consumption for heat energy production etc.;
- energy security parameters: total backup pumps power, total volume of accumulator tanks etc.;
- economic parameters: heat cost, annual release of heat to consumers etc.;
- operational reliability parameters: technical readiness coefficient by time, presence of protective coating on the pipes etc.

The columns of the dataset include *Num* which defines the number of the row, *isCGP* which contains Boolean data defining the type of object (1 for cogeneration plants and 0 for boiler houses). Other columns of the dataset contain common parameters of the objects (e.g. *fuel*, *sumPumpPow* – total power of the pumps installed, *anNOxEm* – annual specific $NO_x$ emissions etc.). Certain columns are object-specific. *numTurb* – number of turbines or *eeBoilCGP* – energy efficiency class of additional boilers at a CGP are used only for CHP objects; *weathReg* – presence of weather-dependent regulation or *avgTempOutBH* – average output temperature at a boiler house are used only for boiler house objects.

The dataset's data is transferred to the input of the block, which implements the selection of columns used for further training. In this case, the block selects all columns except *Num*, which contains the serial numbers of objects. Column selection is done using the "Column selector" tool.

Using a script written in the Python programming language, dataset columns containing unique parameters of objects of a different type are deleted based on the value of the *isCGP* parameter specified in the "Enter Data Manually" block. For example, when selecting the "Boiler Plant" object type, dataset columns containing parameters unique to CHP are deleted. The block itself is a simplified version of the dataset, containing only one value of one parameter. In this case, the boiler plant selection corresponds to the value *isCGP* = 0, CHP – *isCGP* = 1.

The "Clean Missing Data" block deletes dataset rows containing empty values. The cleaned dataset is fed to the "Split Data" block, which divides the dataset rows into data for training the model and data for testing. In this model, 75% of the rows are allocated for training the model. Thus, the distribution of the data rows for training and evaluation of the model is 75%/25%.

## 4. Results & Discussions

Regression using a neural network is used to predict the output values. The neural network approach has advantages over traditional methods in three cases:

- When the problem under consideration, due to its specific features, is not amenable to adequate formalization, since it contains elements of uncertainty that are not formalized by traditional mathematical methods;
- When the problem under consideration is formalized, but at present there is no apparatus for its solution;
- When there is an appropriate mathematical apparatus for the considered, well-formalized problem, but the implementation of calculations with its help on the basis of computing systems does not satisfy the requirements for obtaining solutions in terms of time, size, weight, energy consumption, etc. In such a situation, it is necessary to either simplify the calculation methods, which reduces the quality of solutions, or apply the appropriate neural network approach, provided that it will provide the necessary quality of task performance.

Due to the lack of possibility of a clear, comprehensive formalization of engineering systems, including heat supply systems, it is advisable to choose regression using a neural network.

The parameters of the neural network used in these models are given in Table 2.

Table 2. Neural network parameters

| Parameter | Value |
|---|---|
| Number of hidden layers | 60 |
| Learning rate | 0.005 |
| Number of learning cycles | 60 |
| Primary weights diameter | 0.1 |
| Gradient | 0 |

After training the model using the "Train Model" block and testing it using the "Score Model" block, it is evaluated in the "Evaluate Model" block. The results of the evaluation of this model for boiler rooms are given in Table 3.

Table 3. Model evaluation results for boiler houses

| Parameter | Value |
|---|---|
| Mean absolute error | 0.086892 |
| Mean squared error | 0.110108 |
| Relative absolute error | 0.29382 |
| Relative squared error | 0.07819 |
| Determination coefficient | 0.92181 |

Accordingly, the accuracy of predicting the value of the objective function using the model for boiler rooms is 92.18%.

## 4. Conclusions

The adaptive decision-support model described in the paper can be used as an efficient tool in managing engineering infrastructure reconstruction programs. Its effectiveness in program architecture development, management, and work breakdown structure (WBS) creation is ensured by its adaptability, scalability, and the inclusion of post-processing tools and variable parameter redistribution components. These features enable decision-makers to tailor the model for various types of objects within the system being reconstructed under the program.

The model evaluation results conducted in Microsoft Azure Machine Learning Studio show a high level of precision in objective function prediction.

The model's development approach makes it applicable to the reconstruction of diverse engineering infrastructure systems, including heating, gas supply, electricity supply, and drainage networks, among others.

## Acknowledgement

The research was supported by the National Research Foundation of Ukraine (Grant Agreement No. 2023.03/0131) and partially by the European Union Assistance Instrument for the Fulfilment of Ukraine's Commitments under the Horizon 2020 Framework Programme for Research and Innovation (Research Project No. 0123U102775).